\documentclass[letterpaper,twocolumn,prl,aps,superscriptaddress,floatfix,showpacs]{revtex4-1}

\usepackage{multirow,amssymb,amsbsy,amsmath}
\usepackage{graphicx}
\usepackage{verbatim}
\usepackage{booktabs}
\usepackage{dcolumn}
\makeatletter

\newcommand{\braket}[2]{\langle #1#2\rangle}

\newcommand{\ita}[1]{\textit{#1}}

\usepackage{color}

\begin{document}

\title{Local probe for connectivity and coupling strength in quantum complex networks}

\author{Johannes Nokkala}
\email{jsinok@utu.fi}
\affiliation{Turku Centre for Quantum Physics, Department of Physics and Astronomy,
University of Turku, FI-20014, Turun Yliopisto, Finland}
\author{Sabrina Maniscalco}
\affiliation{Turku Centre for Quantum Physics, Department of Physics and Astronomy,
University of Turku, FI-20014, Turun Yliopisto, Finland}
\affiliation{Centre for Quantum Engineering, Department of Applied Physics, School of Science, Aalto University, P.O. Box 11000, FIN-00076 Aalto, Finland}\author{Jyrki Piilo}
\affiliation{Turku Centre for Quantum Physics, Department of Physics and Astronomy,
University of Turku, FI-20014, Turun Yliopisto, Finland}

\date{\today}

\begin{abstract}
We develop a local probe to estimate the connectivity of complex quantum networks.
Our results show how global properties of different classes of complex networks can be estimated -- in quantitative manner with high accuracy -- by coupling a probe to a single node of the network. Here, our interest is focused on probing the connectivity, i.e. the degree sequence, and the value of the coupling constant within the complex network. The scheme combines results on classical graph theory with the ability to develop quantum probes for networks of quantum harmonic oscillators. Whilst our results are proof-of-principle type, within the emerging field of quantum complex networks they may have potential applications for example to the efficient transfer of quantum information or energy or possibly to shed light on the connection between network structure and dynamics.
\end{abstract}


\maketitle

\section{Introduction}

While the study of classical complex networks has enjoyed considerable interest throughout the last 20 years \cite{BarabasiRMP,Newman,Barabasi17}, the study of interacting quantum systems as quantum complex networks has only recently started to emerge \cite{Bianconi15,Biamonte17}. The topics range from state \cite{quanttransport} and energy transfer \cite{quanttransportenergy} as well as random quantum walks \cite{Faccin13} on such networks to modeling structured finite environments \cite{complexnetworkenvironments} and investigating the possible quantum effects in photosynthesis \cite{quantphoto}. Quantum networks are also important in development of more complicated quantum communication schemes \cite{qcn1,qcn2}. Experimental platforms that could be used to implement the quantum complex networks in the near future include arrays of micromechanical resonators cooled near to their ground state \cite{resonatorarrays}, cold atoms in lattices \cite{coldatoms} and cluster states or networks of bosonic modes \cite{paris1,paris2,paris3}.

Broadly speaking, networks are any systems that can be thought of as being composed of many interacting or otherwise related subsystems or entities. This includes an immense variety of large complex systems such as acquaintance networks \cite{socialnetworks}, the global shipping network \cite{gsn} and food webs in an ecosystem \cite{foodweb1,foodweb2}, but also microscopic ones like metabolic processes in a cell \cite{metabolic1,metabolic2} and light-harvesting complexes \cite{FMOcomplex}. The ability to capture the essential features of so many different systems of interest makes network theory a powerful tool. Much of its power stems from reducing a complicated system into an abstract graph composed of nodes connected by links. This can then be studied independently of what the physical network is and revealing, e.g., important information on mechanisms influencing the construction and evolution of these complex systems. This is expected to hold true even if the constituents of the complex network are quantum physical objects.

An important problem in network theory is the extraction of information about the network when only a small subset of its constituents can be accessed. This has also been considered in the quantum case, and it has been shown that, provided one has suitable prior knowledge of the network, it is possible to determine several of its properties indirectly using a probe system, such as the network state  \cite{probeState}, temperature \cite{qnt1,qnt2}, and coupling strengths between nodes \cite{Daniel2009,qubitchains}. In particular, in the case of full access, the structure of the network can in principle be determined exactly \cite{complexnetworkenvironments,Daniel2011}. The developed theoretical tools are crucial on the one hand for understanding how the structure of a nontrivial quantum environment is encoded in the dynamics of an open quantum system, and on the other hand for identifying and measuring the key properties of different quantum networks.

In this work, we consider the estimation of connectivity given by the number of links, or degree, of each node in the case of a simple and connected abstract graph.  This choice is motivated by the fact that the degree sequence and corresponding distribution is one of the most important and commonly used concepts in characterising complex networks. 
By simple, we mean that between any two nodes there is at most one link and no node has a link with itself, and by connected that any node can be reached from any other by following the links. We also assume that the links are undirected, meaning that the interactions or relations modeled by the links are taken to be symmetric.
Our results are general in the sense that the only assumption one must make about a physical network is that we know the number of nodes within the network and it is possible to perform measurements with results that are in a known relationship with the eigenvalues of the Laplace matrix of the corresponding graph. As an example of this type of system, we use a network of identical quantum harmonic oscillators interacting with spring-like couplings of constant magnitude \cite{complexnetworkenvironments,PalmaScirep}.

Earlier work for quantum networks has been done in the case of networks of spins, based on continuous-measurement-based approach of small networks up to $5$ nodes with uniform or approximately uniform couplings \cite{Kato2014}, as well as for quantum oscillator networks where the mutual information between a node and the rest of the network was shown to be characteristic of the topology when the network is at or near its ground state \cite{infosharingnetworks}. In contrast, our approach can in principle be applied to any kind of classical or quantum networks as long as the Laplace eigenvalues can be extracted. In practice, the amount of available computational power will limit the size of the networks.

Our main result is that it is indeed possible to obtain accurate estimate for the degree sequence of the network by using only a single probe that is coupled to one of the nodes of the complex network. This result is based on
exploiting known mathematical relations between the Laplace eigenvalues and the connectivity, and using the possibilities that quantum probing provides. The numerical evidence shows that the scheme works very well for different classes of network structures and is robust to small errors in the probed quantities.  We also consider the case where the coupling strength in an oscillator network is uniform but a priori unknown. It turns out that for some classes of networks the coupling strength can always be correctly deduced, and numerical evidence suggests that the estimation succeeds with high probability in the general case.
 
For the sake of simplicity, we show first - in terms of classical graph theory -  how the degree sequence of complex networks can be estimated once the eigenvalues of the Laplace matrix are known. After this, we turn our attention to quantum networks and develop a scheme to probe locally these eigenvalues and the corresponding eigenfrequencies within the network of quantum harmonic oscillators.

\section{Connectivity estimation} 

\begin{figure}[t]
                \includegraphics[trim=3.8cm 0.25cm 2.5cm 0.5cm,clip=true,width=0.45\textwidth]{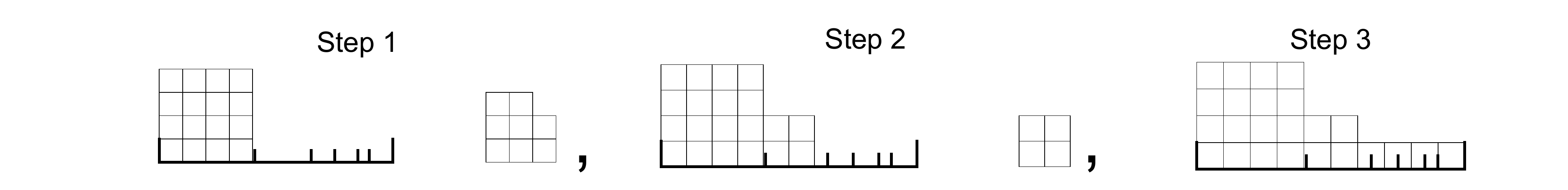}
            \caption{\label{fig0}
(Color online) A visualization of the constraints imposed on the degree sequence by equations $(2)$, $(3)$ and $(4)$. The length of the line is given by Eq. $(2)$. The number of boxes coincides with the sum of squared degrees given by Eq. $(3)$, and the ticks are given by the bounds appearing in Eq. $(4)$. By filling the line with squares such that all boxes are used, there are as many squares as there are slots, and each square will not exceed its slot, one will find an integer sequence satisfying simultaneously the three equations. Here this is done in three steps for a small graph to illustrate the problem. The bounds on mimimum and maximum degree are not shown.  
}
      \end{figure}

Once the nodes of a simple and connected graph have been labeled, its structure may be encoded into a matrix in many ways. In particular, the Laplace matrix $\textbf{L}$ of the graph has elements 
\begin{equation}
\label{eq.1}
L_{ij}=\delta_{ij}d_i - (1 - \delta_{ij})l_{ij},
\end{equation}
\noindent
where $d_i$ is the degree of node $i$ and $l_{ij}=1$ if there is a link between nodes $i$ and $j$ and $0$ otherwise; notice that $l_{ij}=l_{ji}$. Given the eigenvalues $\lambda_i$ of the Laplace matrix, the objective is to estimate the degrees $d_i$. This can be done by combining several results from spectral graph theory, which studies the relationship between graphs and the eigenvalues of their matrices.
In addition to bounds on minimum and maximum degree by eigenvalues $\lambda_i$, the following relations must be fulfilled \cite{spectraofgraphs,grone}
\begin{align}
\text{Tr}\textbf{L}=\sum\nolimits_i^N d_i&=\sum\nolimits_i^N \lambda_i, \\
\text{Tr}\textbf{L}^2-\text{Tr}\textbf{L}= \sum\nolimits_i^N d_i^2&=\sum\nolimits_i^N (\lambda_i^2-\lambda_i), \\
1+\sum\nolimits_i^{m<N} d_i &\leqslant \sum\nolimits_i^{m<N} \lambda_i.
\end{align}
The above restrictions are illustrated using a small example in Fig. ~\ref{fig0}. We use a method to construct sequences $\mathbf{d}'$ of $N$ positive integers that satisfies simultaneously the degree bounds and restrictions  $(2)$, $(3)$ and $(4)$, and call them solutions. 

There are special cases where there is only one solution, and consequently the degree sequence $\mathbf{d}$ is unambigiously determined by the eigenvalues. It is straightforward to see that these include a simple chain, completely connected network and all regular graphs, i.e. graphs for which all degrees coincide. For the first two this follows from the fact that the squared sum in Eq.~$(3)$ attains its minimimum or maximum value for a given number of nodes, while a direct application of Cauchy-Schwarz inequality shows that only regular graphs have equality in $(\sum_i^N d_i)^2 \leqslant  N \sum_i^N d_i^2$. There is also an important class of graphs called threshold graphs \cite{trg} that are uniquely determined by their degree sequence and their degree sequence is in turn determined by the Laplace eigenvalues, however the eigenvalues will typically be degenerate. 

More generally, Eqs.~$(2)$ and $(3)$ for a given $N$ fix the mean and variance of the bounded solutions and Eq.~$(4)$ further refines them by ruling out cases where deviations from $\mathbf{d}$ are bunched together. For any solution, the deviations must cancel out because the correct sum of degrees is enforced; similarly also deviations between any element-wise squared solution and the element-wise squared $\mathbf{d}$ must cancel out. Since the possible values of degrees are integers, the number of solutions is finite. Given enough computational power and time, all of them can be found; this is feasible on a normal tabletop computer when the number of solutions is in the lower end of hundreds of thousands, limiting $N$ to tens of nodes. We stress that different classes of networks will have different scaling between $N$ and the number of solutions. 

To find the solutions, we consider Eq. $(3)$ as an integer partitioning problem, where the sum of squared elements must be partitioned into $N$ integers. The allowed integers are square numbers with bounds determined from the eigenvalues. Taking the element-wise squareroots of each found partition and filtering the results according to equations $(2)$  and $(4)$ will provide the solutions. Alternatively, one could start from Eq. $(2)$ and then filter but we found that this is more wasteful and consequently uses more memory and computation time.  

We tested our estimation scheme on Erd\H{o}s-R\'{e}nyi random graphs \cite{ER}, Barab\'{a}si-Albert graphs \cite{BA}, Watts-Strogatz graphs  \cite{WS} and tree graphs. An Erd\H{o}s-R\'{e}nyi random graph refers to either of two closely related models of generating random graphs. In both, the number of nodes is fixed. Using the so-called $G(N,L)$ model, one chooses uniformly among all possible graphs with $N$ nodes and exactly $L$ links, while using the $G(N,p)$ model, one starts from a completely connected graph and includes each link in the final graph with probability $p$. Here we use the former model unless otherwise stated. A Barab\'{a}si-Albert random graph $G(N,K)$ is constructed starting from a cyclic graph of three nodes and iteratively adding a new node with $K$ links until the graph has $N$ nodes, connecting the new links randomly but favoring nodes with higher degree. It can be shown that graphs constructed like this have a degree distribution that follows a powerlaw. Watts-Strogatz graphs $G(N,k,p)$ are constructed by starting from a circular graph where each node is connected to up to $k$-th nearest neighbors. Then each link is rewired with probability $p$, creating a graph with small world properties. Finally, a tree of $N$ nodes is any connected graph with exactly $N-1$ links; this gives them the property that they have no cycles, i.e. closed walks without repetitions of links or nodes other than the starting and ending node.

As a figure of merit of a solution $\mathbf{d}'$ we chose the $\ell_1$ distance from $\mathbf{d}$ normalized by the total degree of the graph, i.e. 
\begin{equation}
f(\mathbf{d}')=\Vert \mathbf{d}-\mathbf{d}' \Vert_1/\Vert \mathbf{d}\Vert_1= \sum\nolimits_i^N |d_i-d'_i|/|d_i|.
\end{equation}

This choice is motivated by the fact that this quantity can be interpreted as the average deviation from the real degree per link. We found that, for all considered cases, $f(\mathbf{d}')<1/2$. By choosing as final estimate the solution that has the smallest $\ell_1$ distance from the mean of solutions it is possible to single out a solution particularly close to $\mathbf{d}$, since the deviations, that must cancel out for any particular solution as explained previously, will then be partly averaged out. By mean of solutions, we indicate the sequence where each element is the corresponding mean degree calculated from all solutions. On the other hand, the set of all solutions always contains $\mathbf{d}$ while the estimate is typically not a perfect match.

\begin{figure*}[t]
                \includegraphics[trim=0cm 0cm 0cm 0cm,clip=true,width=0.95\textwidth]{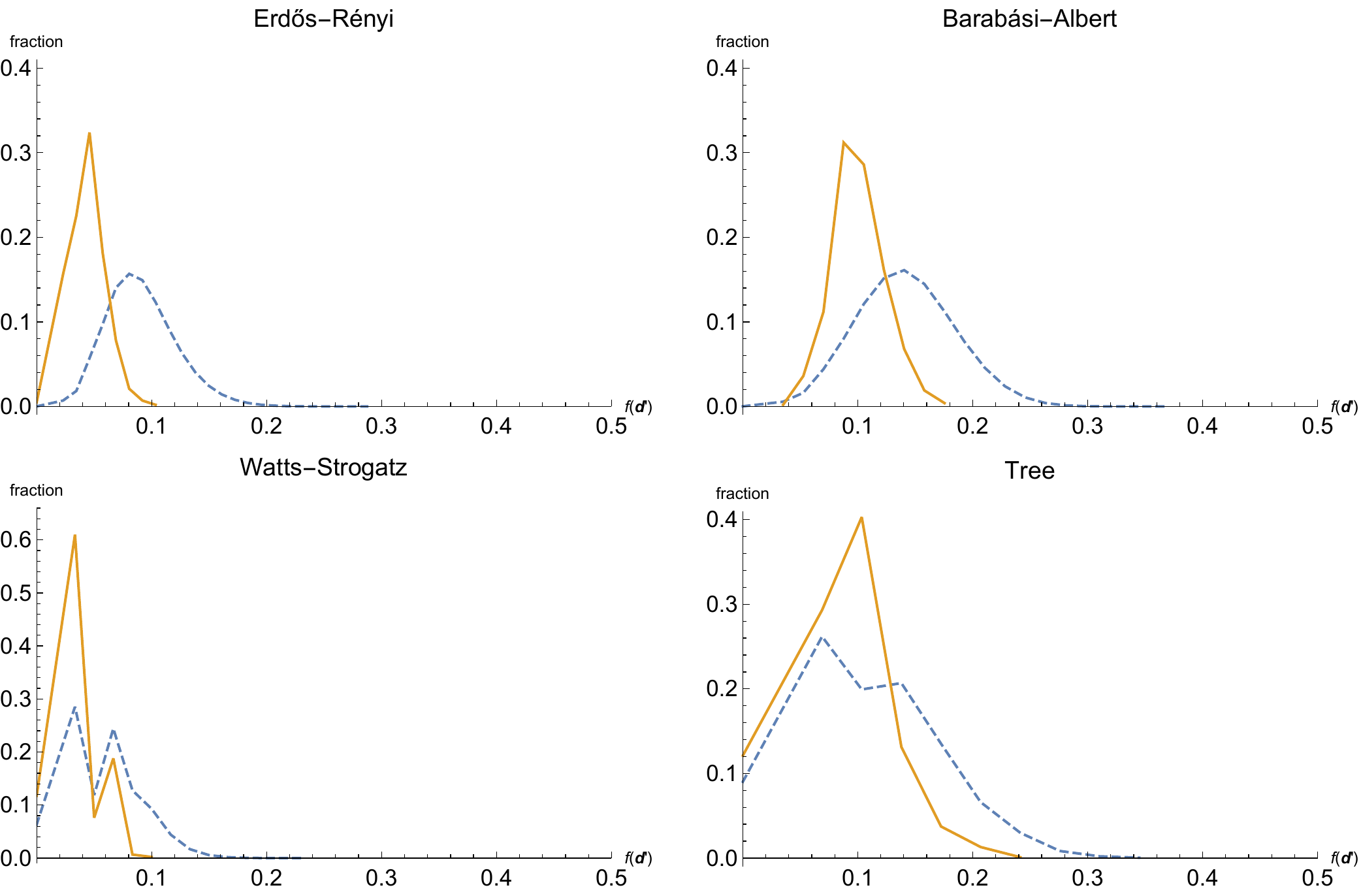}
            \caption{\label{fig1}
(Color online) Solutions $\mathbf{d}'$ compared by their distance from the real degree sequence $\mathbf{d}$ as quantified by $f(\mathbf{d}')=\lVert \mathbf{d}-\mathbf{d}' \lVert_1/\Vert \mathbf{d}\Vert_1$ for different types of random graphs, each having $N=30$ nodes. The dashed line corresponds to all solutions satisfying the constraints and the solid line to estimates acquired by choosing the solution closest to their mean. Refer to main text for details. All results are averaged over $1000$ realizations of each type of graph.
}
      \end{figure*}

The results, averaged over 1000 realizations with network size fixed to $N=30$, are shown in Fig.~\ref{fig1}. Besides the parameter values considered here, we have also checked other values and found similar results. For Erd\H{o}s-R\'{e}nyi random graph, we used $L=87$. This would be the expected number of links for $G(N,p)$ of same size with $p=1/5$. For Barab\'{a}si-Albert graph, we used $K=2$. In the latter case the estimation performs worst, and in particular none of the estimates coincided with the real degree sequence. This is caused by the high variance of $\mathbf{d}$ for this class of random graphs: higher variance allows the solutions to deviate more from $\mathbf{d}$ and consequently the estimation is less accurate.

Compared to the other two graphs which had typically thousands of solutions, Watts-Strogatz graphs and trees had much less solutions, with the former having tens and the latter only a handful with the used parameter values. Consequently a significant fraction of estimates were a perfect match with $\mathbf{d}$. The plots are not smooth, indicating that certain values are much less likely than others, a feature not present for the other two graphs. For the former, we used $k=2$ and $p=0.2$. Unlike for the other graphs, more than half of the solutions had the same distance from $\mathbf{d}$. We believe this to be because this class of random graphs had the smallest variance of $\mathbf{d}$ since they are generated from regular graphs. Trees had the biggest fraction of perfect matches out of all graphs, but this is mostly because the number of solutions was so small to begin with. This is essentially caused by any tree having the smallest possible number of links for a given number of nodes, greatly restricting also the solutions.

While $\mathbf{d}'$ close to mean solution are alike, the outliers are different from both them and $\mathbf{d}$. This is because there are many relatively smooth sequences that satisfy the constraints, but only a few jagged ones that pass. Indeed, the estimation works poorly on graphs with jagged degree sequences since the majority of solutions will be much smoother. We stress that choosing an outlier and realizing it as a network will in general not yield the same solutions since degree bounds and restrictions imposed by Eq.~$(4)$ can change even between different realizations of a fixed $\mathbf{d}$.

\section{Application to quantum networks}

To exploit the previous results for quantum probing and networks, we consider networks of uniformly coupled quantum harmonic oscillators \cite{complexnetworkenvironments}. We will use units as referred to an arbitrary (but fixed) frequency unit and give coupling strengths, times and temperatures in terms of this unit. We will also set $\hbar=1$ and $k_B=1$. The network is composed of $N$ unit mass quantum harmonic oscillators coupled by springs, each having the same bare frequency $\omega_0$. The couplings between network oscillators are assumed to be uniform with the strength given by $g$. We can express the network Hamiltonian in a compact way as 
\begin{equation}
H_{E}=\textbf{p}^{T}\textbf{p}/2+\textbf{q}^{T}(\omega_0^2\textbf{I}+g\textbf{L})\textbf{q}/2, 
\end{equation}
\noindent where $\textbf{p}=\left\lbrace p_{1}, p_{2}, ..., p_{N}\right\rbrace ^T  $ and $\textbf{q}=\left\lbrace q_{1}, q_{2}, ..., q_{N}\right\rbrace ^T  $ are the vectors of momentum and position operators, $\textbf{I}$ is the identity matrix and $\textbf{L}$ is the Laplace matrix of the underlying graph. We will assume that $g$ and $N$ are known, but make no assumptions on $\textbf{L}$. Since the row sums of any Laplace matrix are zero, the eigenvalues $\lambda_i$ are non-negative. This, together with a positive coupling constant $g$, ensures the positivity of Hamiltonian $H_E$.

Since the network Hamiltonian is quadratic in position and momentum operators for any configuration given by $\textbf{L}$, it can be diagonalized with an orthogonal transformation. This allows us to move into an equivalent picture of noninteracting eigenmodes of the network. In this picture, $H_E=\sum\nolimits_{i=1}^N (P_{i}^2+\Omega_{i}^2Q_{i}^2)/2$, where $P_{i}$ and $Q_{i}$ are the position and momentum operators of the network eigenmodes and $\Omega_{i}$ are their frequencies, related to the eigenvalues $\lambda_i$ of the Laplace matrix $\textbf{L}$ as
\begin{equation}
\label{eq.2}
\lambda_i=(\Omega_{i}^2-\omega_0^2)/g.
\end{equation}
This is the key equation which allows us to use the previously described estimation procedure for the degree sequence.
In other words, if we can probe the eigenfrequencies $\Omega_i$ of the network, this gives us direct information about the eigenvalues of the Laplace matrix and therefore a way to estimate the connectivity of the network.
It is also worth mentioning that, since $\omega_0$ coincides with the smallest eigenfrequency, it is not necessary to know it beforehand.

Assuming that the network is in a thermal state of known temperature $T$, the detection of eigenfrequencies can be done by measuring the mean excitations $\braket{n(t)}{}$ of a bosonic probe weakly coupled to a node in the network and doing a frequency sweep across the range that covers the spectrum \cite{complexnetworkenvironments}. The probe is assumed to be a quantum harmonic oscillator with the Hamiltonian $H_S=(p_{S}^2+\omega_{S}^2q_{S}^2)/2$, where $p_{S}$ and $q_{S}$ are its momentum and position operators and $\omega_{S}$ is its frequency, while the interaction Hamiltonian is of the form $H_I=-kq_Sq_j$, where $k$ is the strength of the coupling and $q_j$ is the position operator of the node interacting with the probe. By fixing the states of the probe and the network, the reduced dynamics of the probe can be determined exactly by diagonalizing the total Hamiltonian, solving the Heisenberg equations of motion for the decoupled oscillators, and returning to old operators. While here we fix the state of the probe and the network to be vacuum and thermal state of temperature $T$, respectively, the accuracy is largely insensitive to the state of the probe as long as there is an energy difference between the probe and the network \cite{complexnetworkenvironments}.

\begin{figure}[t]
                \includegraphics[trim=0cm 0cm 0cm 0cm,clip=true,width=0.45\textwidth]{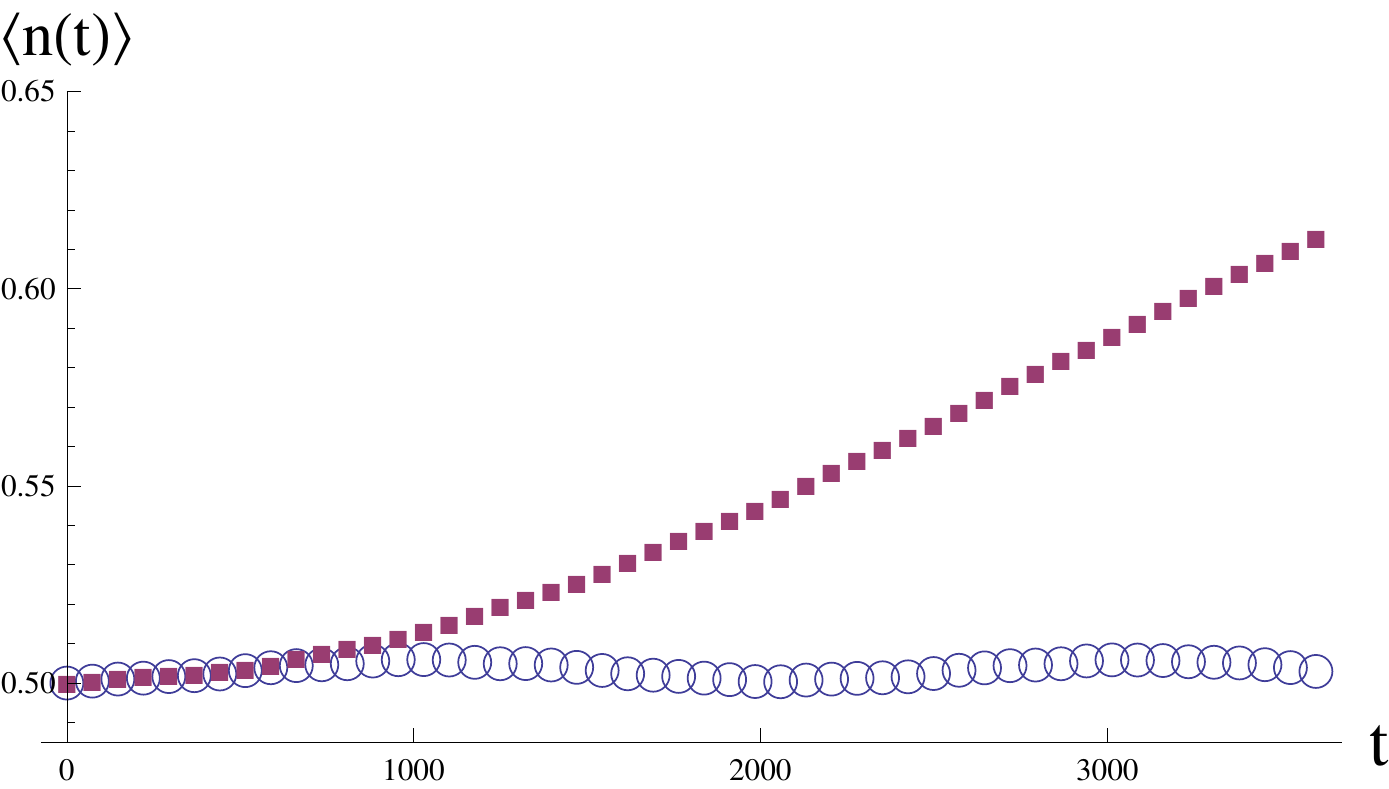}
            \caption{\label{fig2}
(Color online) Evolution of mean excitations $\braket{n(t)}{}$ for a probe system weakly coupled to a single node in an Erd\H{o}s-R\'{e}nyi network of $40$ oscillators and $80$ couplings with bare frequency $\omega_0=0.2$ and coupling constant $g=0.1$, with probe frequency coinciding with an eigenfrequency of the network (squares) and just a little above 1 \% off (circles). The initial state of the probe and the chain were vacuum and thermal state with $T=0.3$, respectively, while coupling strength between the probe and the chain was $k=0.0025$. The clear difference in the dynamics for longer interaction times makes the detection of eigenfrequencies possible. Once detected, the eigenfrequencies can be used to determine the Laplace eigenvalues.  
}
      \end{figure}

When coupled strongly to the network, the probe will exchange energy with all eigenmodes and the reduced dynamics depends on the structure of the network in a complex way. On the other hand, with a sufficiently weak coupling the interaction becomes limited to only the few closest modes in the vicinity of system frequency $\omega_{S}$, and this makes the reduced dynamics very sensitive to the resonance condition in the sense that when $\omega_{S}$ matches an eigenfrequency, a significantly larger amount of energy can flow between the network and the probe before finite size effects cause the flow to be reversed. An example is shown in Fig. \ref{fig2}, which demonstrates that even a small difference in frequencies can lead to a very different value of $\braket{n(t)}{}$, for sufficiently long interaction times and a weak coupling, provided that there is an energy difference between the probe and the network. While this behaviour is universal to finite networks, the number of nodes $N$ is assumed to be known in the probing protocol because otherwise one does not know when all eigenfrequencies have been found.

The probe must interact with an eigenmode to detect its frequency. The spectrum should also be nondegenerate because any degenerate eigenfrequencies are interpreted as a single frequency. This is typically the case, and it can be seen by considering the oscillators in terms of the eigenmodes: any $q_i$ can be expressed as a weighted sum of eigenmode position operators where the weights are given by the elements of the $i$th eigenvector of the matrix $(\omega_0^2\textbf{I}+g\textbf{L})/2$. For a generic $\textbf{L}$, all eigenvalues are distinct and the eigenvectors will not have zero elements, which means that the probe will interact with and resolve all eigenmodes from any node.

In the non-ideal case, there might be some errors in the values of eigenfrequencies or the coupling strengths might be only approximately uniform. We checked the robustness against both for all four classes of networks. For all of them, $1$ \% unbiased error in either eigenfrequencies or coupling strengths will typically not cause any errors in the detected sum of degrees while perturbing the probed sum of squared degrees, degree bounds, and bounds on partial sums very little if at all. With larger errors, the worst case accuracy of results averaged over many realizations deteoriates slowly, but the differences between individual realizations grows. We also found that the number of solutions had a large impact on the robustness of the best case accuracy, as this was affected very quickly for trees and Watts-Strogatz networks while the other two classes of networks were much more resilient. Sometimes the affected bounds on partial sums did not provide any solutions at all for trees or Watts-Strogatz networks, in which case we considered the accuracy of solutions without this restriction.

In the case of nonuniform coupling strengths, the eigenvalues of a weighted Laplace matrix $\textbf{L}$ can be recovered from $\Omega_i^2-\omega_0^2$ . Now the off-diagonal elements of $\textbf{L}$ are the coupling strengths between the oscillators and the diagonal has the sums of coupling strengths to each oscillator. While other restrictions still apply as before, the eigenfrequencies only upper bound the sum of the squares of diagonal elements of $\textbf{L}$ and conversely, their variance can only be bounded from above, reducing the accuracy of the estimation considerably. The number of possible solutions can still be finite if the coupling strengths in the network are divisible by the same number, for instance if there is a weakest coupling and others are its integer multiples.

\section{Estimation of an unknown coupling constant}

If the coupling strengths are known to be uniform but the value of the coupling constant is not known, one can estimate it from the probed eigenfrequencies using the relation $g \lambda _i=\Omega_i^2-\omega_0^2$ obtained from Eq. $(7)$. The estimation procedure uses general properties of the eigenvalues $\lambda_i$ of an unweighted connected graph. We stress that the success or failure of the estimation depends only on the structure of the graph, rather than on a particular value of $g$. As will be seen below, for generic degree sequences it succeeds.

\begin{figure}[t]
                \includegraphics[trim=0cm 0cm 0cm 0cm,clip=true,width=0.45\textwidth]{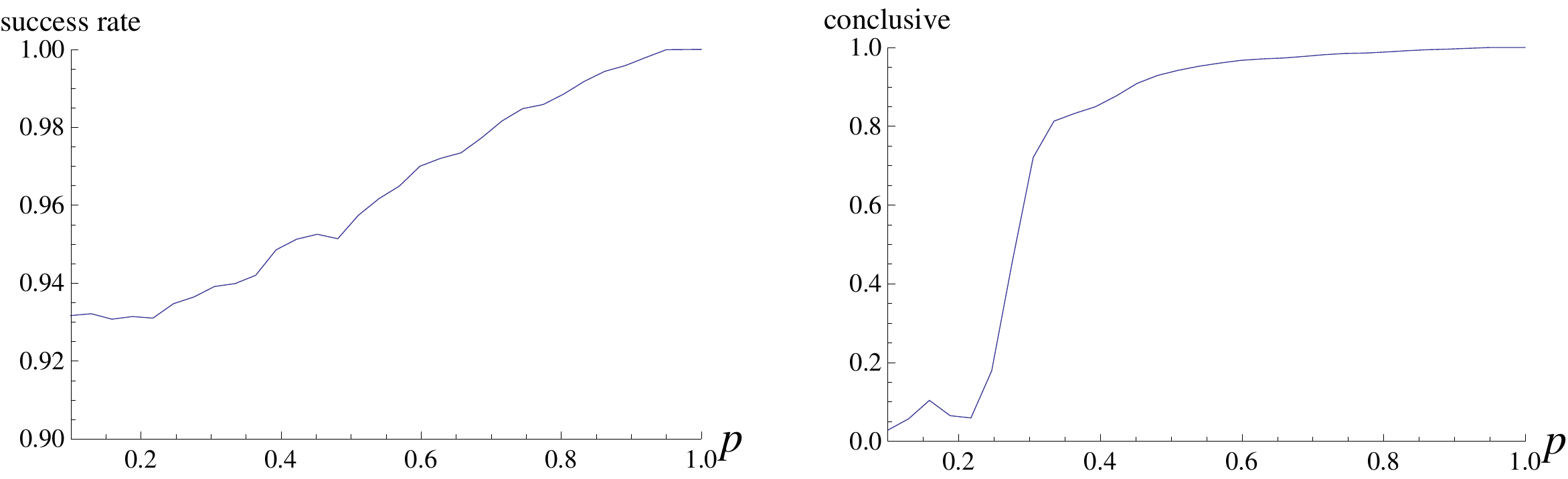}
            \caption{\label{fig3}
(Color online) Estimation of an unknown coupling constant $g$ for the Erd\H{o}s-R\'{e}nyi random network, compared with connection probability $p$. The estimation procedure produces a finite list of possible values for the coupling constant, of which the largest is selected as the estimate. The fraction of cases where the estimate coincides with $g$ is shown on the left. On the right, the fraction of cases where $g$ is the only value produced by the estimation is shown. Refer to main text for details. All results are averaged over $50000$ realizations of the network.
}
      \end{figure}

Because the graph is connected and simple, we know that $2(N-1)\leq \sum_i^N d_i \leq N(N-1)$. Since $\sum_i^N d_i = \sum_i^N \lambda_i$, this leads to $\frac{\sum_i^N(\Omega_i^2-\omega_0^2)}{N(N-1)}\leq g \leq \frac{\sum_i^N(\Omega_i^2-\omega_0^2)}{2(N-1)}$. We can reduce this range to a finite set of values by demanding that both $\sum_i^N d_i$ and $\sum_i^N d_i^2$ are even integers, as they must be for a connected graph. This set can be further refined by using results related to regular graphs and the largest eigenvalue $\lambda_N$. As mentioned before, for any connected graph $N \sum_i^N d_i^2-(\sum_i^N d_i)^2 \geq 0$ with equality iff the graph is regular. This property can be violated for values of the coupling constant larger than $g$, which can be used to rule them out. On the other hand, values smaller than $g$ can violate the property $\lambda_N \leq N$ \cite{largesteig}. Typically several values pass these tests, however as we will argue below, they are not equally likely to be correct. 

Clearly, if some $g'$ satisfies the condition that both $\sum_i^N d_i$ and $\sum_i^N d_i^2$ are even, then so does any $g'/x$ where $x=2,3,4,...$. This suggests that $g$ is more likely to be among the larger values satisfying the constraints. In fact, for trees and regular networks, the largest possible value coincides with $g$. In the former case this follows directly from the fact that the sum of degrees attains its minimum value, and hence any $g'>g$ will violate the assumption that the network is connected. In the latter case this can be seen by letting $g'=ag$ and noticing that then $N \sum_i^N d_i^2-(\sum_i^N d_i)^2=\frac{N^2\Delta(1-a)}{a^2}<0$ for all $a>1$, where $\Delta$ is the constant degree of the network.

More generally, for some $g'>g$ to lead to even sum of degrees and squared degrees, it has to be the case that $\sum_i^N (d_i^2+d_i)d_i^2/D'^2$ is even, where $D'<\sum_i^N d_i$ is the wrong sum of degrees corresponding to $g'$. While such a $g'$ might still be ruled out by the other constraints, this implies that without  prior knowledge of the structure of the network, $g$ can be determined unambigiously only when no other value passes the tests. We studied how well the estimation works in the case of the Erd\H{o}s-R\'{e}nyi random network as a function of connection probability $p$, as shown in Fig \ref{fig3} -- unlike in the previous section, here we use the $G(N,p)$ model since for prime values of the total degree the estimation almost always succeeds, and consequently the $G(N,L)$ model leads to a discontinuous plot. The results confirm that the largest value coincides with $g$ with high probability, success rate improving for larger values of $p$. Also shown is the fraction of conclusive cases, i.e. when $g$ is the only possible value. The curve shows an interesting behaviour, with a sudden transient from most cases being inconclusive to most being conclusive, between $p=0.2$ and $p=0.4$. This is essentially because then $\lambda_N>N/2$, which will rule out any $g'\leq g/2$. This does not guarantee conclusiveness since some $g>g'>g/2$ might still pass, but this requires special values of  $\sum_i^N d_i$ and $\sum_i^N d_i^2$.

\section{Discussion and conclusion}

Connectivity is an important structural property of complex networks. We considered simple connected graphs and showed how connectivity can be estimated from the eigenvalues of the Laplace matrix. Our estimation scheme is applicable to any network, quantum or classical, amenable to the extraction of Laplace eigenvalues from measurement results. While the accuracy is best for networks with a degree sequence having small variance, the estimation performs well also for, e.g., networks where the degrees follow a powerlaw. In practice, a network can be too large for completing the entire estimation procedure in a reasonable amount of time, however since the mean and variance of connectivity can be easily and exactly determined from the eigenvalues, the connectivity of these networks can still be classified accordingly. 

We applied our results to networks of identical uniformly coupled quantum harmonic oscillators and showed how not only the connectivity but also the uniform coupling strength can be estimated with local probing of any of the oscillators in the network, making such quantum networks universally suited to the extraction of global properties from locally extractable information.

In this work, we have demonstrated how even in the quantum case, graph theory can be highly useful in eludicating the properties of coupled many-body systems. While here we used information extractable from a quantum network with minimal access, it would be interesting to study the case where a small subset of nodes could be accessed, or investigate how knowing also some of the eigenvectors of the graph could be exploited.

\begin{acknowledgments}
The authors acknowledge financial support from the Horizon 2020 EU collaborative projects QuProCS (Grant Agreenement No. 641277). J. N. acknowledges the Wihuri foundation for financing his graduate studies.   
\end{acknowledgments}

\end{document}